\documentclass{cpbtex}
\usepackage{array}
\usepackage{graphicx}
\usepackage{threeparttable}
\usepackage{amsthm}
\theoremstyle{definition}
\newtheorem{theorem}{\emph{\textbf{Theorem}}}
\newtheorem{lemma}{\emph{\textbf{Lemma}}}

\begin{document}

\title{Quantum search for unknown number of target items hybridizing the fixed-point method with the trail-and-error method\thanks{Project supported by the National Natural Science Foundation of China (Grants No.~11504430 and No.~61502526) and the National Basic Research Program of China (Grant No.~2013CB338002).}}


\author{Tan Li$^{1,2}$,  Shuo Zhang$^{1,2}$,  Xiang-Qun Fu$^{1,2}$,   Xiang Wang$^{1,2}$, \\  Yang Wang$^{1,2}$,  Jie Lin$^{1,2}$,  and  Wan-Su Bao$^{1,2}$ \thanks{Corresponding author. E-mail:~bws@qiclab.cn}\\
$^{1}${Henan Key Laboratory of Quantum Information and Cryptography,} \\
{PLA SSF IEU, Zhengzhou, Henan 450001, China} \\  
$^{2}${Synergetic Innovation Center of Quantum Information and Quantum Physics,}\\
{University of Science and Technology of China, Hefei, Anhui 230026, China}} 

\date{\today}
\maketitle

\begin{abstract}
For the unsorted database quantum search with the unknown fraction $\lambda$ of target items, there are mainly two kinds of methods, i.e., fixed-point or trail-and-error.
(i) In terms of the fixed-point method,
Yoder et al. [Phys. Rev. Lett. 113, 210501 (2014)] claimed that the quadratic speedup over classical algorithms
has been achieved.
However, in this paper, we point out that this is not the case, because the query complexity of Yoder's algorithm
is actually in $O(1/\sqrt{\lambda_0})$ rather than $O(1/\sqrt{\lambda})$, where $\lambda_0$ is a known lower bound of $\lambda$.
(ii) In terms of the trail-and-error method,
currently the algorithm without randomness
has to take more than 1 times queries or iterations
than the algorithm with randomly selected parameters.
For the above problems, we provide the first hybrid quantum search algorithm based on the fixed-point and trail-and-error methods, where
the matched multiphase Grover operations are
trialed multiple times and the number of iterations increases exponentially along with the number of trials.
The upper bound of expected queries as well as the optimal parameters are derived.
Compared with Yoder's algorithm, the query complexity of our algorithm indeed achieves the optimal scaling in $\lambda$ for quantum search, which reconfirms the practicality of the fixed-point method.
In addition, our algorithm also does not contain randomness,
and compared with the existing deterministic algorithm, the query complexity
can be reduced by about 1/3.
Our work provides an new idea for the research on fixed-point and trial-and-error quantum search.
\end{abstract}



\section{Introduction}
For the unsorted database search, the famous Grover algorithm \cite{Grover1996,Grover1997}
achieves a quadratic speedup over classical algorithms.
Afterwards many analyses, generalizations and variants have been investigated \cite{Biron1999,Long1999,Long2001,Long2002,Li2007,Zhang2011,Sun2012,Li2018a,Toyama2008,Toyama2009,Toyama2013,Toyama2019,Zhong2009,Giri2017,Zhang2018,Mehri-Dehnavi2018,Byrnes2018,Ma2017,Li2014,Li2018,Li2019,Pan2019,Pan2019a}. Grover's algorithm has been proven optimal \cite{Bennett1997,Boyer1998,Zalka1999,Nielson2000,Grover2005a}.
However, the Grover algorithm cannot apply to the case where the fraction $\lambda$ of target
items is completely unknown expect for a lower bound, because iterating too much will pass by the target states, that is the so-called souffl\'e problem \cite{Brassard1997}.

For the case of unknown $\lambda$, there are mainly two quantum search methods.
One is the fixed-point method \cite{Grover2005,Yoder2014}.
In Ref.~\cite{Grover2005}, the final state
gradually converges to the target states along with the
iterations, avoiding ``overcooking'' the state. 
However, the algorithm loses the valuable quantum speedup of quantum search \cite{Yoder2014}, which has also been proven asymptotically optimal \cite{Chakraborty2005,Tulsi2006}.

Fortunately, Yoder et al. \cite{Yoder2014} creatively reduces the ``fixed point'' to a bounded region of the target states,
and claimed that the quadratic speedup over classical search
is maintained consequently, which has got many recognitions \cite{Bhole2016,Gurnani2017,Dalzell2017}.
For example, in Ref.~\cite{Dalzell2017}, Dalzell et al. pointed out that:
``it (Yoder's algorithm) produces $\left|E\right\rangle $ (the uniform superposition over all the target states) with fidelity at least $\sqrt{1-\delta^{2}}$ in only $O(\ln(1/\delta)/\sqrt{\omega})$ time, thus exhibiting the quadratic speedup'', where $\omega$ is a known lower bound of the unknown $\lambda$. Note that, for consistency, in the following, $\lambda_{0}$ instead of $\omega$ will be used to denote the lower bound of $\lambda$.

However, as stated in Ref.~\cite{Dalzell2017},
the complexity in the quantum query model \cite{Qiu2018,Cai2018}
of Yoder's algorithm is in $O(1/\sqrt{\lambda_{0}})$ rather than $O(1/\sqrt{\lambda})$, which is dependent on the lower bound $\lambda_{0}$ rather than $\lambda$. For example, if $\lambda_0=1/N$, while $\lambda =1/\sqrt{N}$, then $O(1/\sqrt{\lambda_0})=O(\sqrt{N})$, while $O(1/\sqrt{\lambda})=O(\sqrt[4]{N})$.
As another example, if $\lambda_{0}=1/N$, while $\lambda=(N-1)/N$, then $O(1/\sqrt{\lambda_{0}})=O(\sqrt{N})$, while $O(1/\sqrt{\lambda})=O(1)$.
Therefore, we confirm that the order of query complexity of Yoder's algorithm is in fact not really optimal (for a detailed complexity analysis of Yoder's algorithm see Section~2).

Another method for the case of unknown $\lambda$ is the trial-and-error method \cite{Boyer1998,Younes2008,Younes2013,Okamoto2001}.
In this regard, the original work was proposed by Boyer et al. \cite{Boyer1998}, which trials Grover's algorithm multiple times, where the number of iterations is randomly selected from an exponentially increasing interval along with the number of trials. This randomness enables
the average success probability at each trial is \emph{always} at least 1/4 after the algorithm reaches the critical stage,
and a target item can thus be found with an expected
number of Grover iterations
no more than $4/\sqrt{\lambda}$,
but the algorithm is therefore referred to as a randomized application of Grover's algorithm.
Based on this randomized trial-and-error method, replacing the internal Grover's algorithm \cite{Grover1996} by the partial diffusion algorithm \cite{Younes2004} or fixed-phase algorithm \cite{Younes2013}, Younes et al. proposed two different variants \cite {Younes2008,Younes2013}, but
the number of iterations was not reduced  (for details see Table 1 in Section 4).

In order to remove the randomness of trial-and-error algorithms,
Okamoto et al. \cite{Okamoto2001} directly set the number of Grover iterations exponentially increasing along with the trials and thus obtained a simpler deterministic  trial-and-error algorithm,
where the success probability at each trial can \emph{often} (but not always) be no less than 3/4.
This allows the algorithm to find a target state also in the optimal scaling of quantum search.
But this also results in more than 1 times iterations or queries
than Boyer's randomized algorithm.

In this paper, we expect to design the first quantum search algorithm that hybridizes the fixed-point method and the trial-and-error method for the case of unknown $\lambda$, to confirm that the fixed-point method can also actually achieve the real optimal query complexity by selecting the number of iterations under the trial-and-error method,
and confirm that the number of queries
of the deterministic trial-and-error quantum search algorithm can be
closer to (and even the number of iterations can be fewer than)
the randomized versions.

The paper is organized as follows. Section~2
provides an introduction of Yoder's fixed-point quantum search algorithm as well as an analysis of the query complexity. Section~3
describes our hybrid quantum search algorithm of the fixed-point method and the trial-and-error method. Section~4
discusses the comparisons between the existing algorithms and the algorithm in this paper, which also gives a brief conclusion.

\section{Yoder's algorithm and its query complexity \label{sec:Yoder-algorithm-revisited}}
Yoder's algorithm \cite{Yoder2014}
aims to overcome the loss of quadratic speedup in the original fixed-point quantum search \cite{Grover2005} and avoid the souffl\'e problem 
in the original Grover algorithm \cite{Grover1996}.

The initial state of Yoder's algorithm is prepared by applying operator $A$ to the state $\left|0\right\rangle $, which can be written as
\begin{equation}
\left|\psi\right\rangle =A\left|0\right\rangle =\sqrt{\lambda}\left|\alpha\right\rangle +\sqrt{1-\lambda}\left|\beta\right\rangle ,
\end{equation}
where $A=H$ is the Hadamard transform, $\left|\alpha\right\rangle $ ($\left|\beta\right\rangle $) represents the equal superposition of all target (nontarget) states, i.e.,
\begin{align}
\left|\alpha\right\rangle &  =  \frac{1}{\sqrt{M}}\sum\limits _{x\in f^{-1}\left(1\right)}\left|x\right\rangle ,\label{eq:target_state}\\
\left|\beta\right\rangle &  =  \frac{1}{\sqrt{N-M}}\sum\limits _{x\in f^{-1}\left(0\right)}\left|x\right\rangle ,\label{eq:nontarget_state}
\end{align}
where $\lambda=M/N$ is the fraction of target items, $M$ is the number of target items in the database of size $N$.

Yoder's algorithm performs on $\left|\psi\right\rangle $ the following sequence of
matched-multiphase Grover operations (referred to as Yoder's sequence)
\begin{equation}
G\left(\phi_{l},\varphi_{l}\right)G\left(\phi_{l-1},\varphi_{l-1}\right)\cdots G\left(\phi_{1},\varphi_{1}\right).\label{eq:sequence-of-operations}
\end{equation}
Here $l$  denotes the number of iterations, $G\left(\phi,\varphi\right)  = -AS_{0}^{\phi}A^{\dagger}S_{f}^{\varphi}$ is the generalized Grover iteration \cite{Brassard2002},
$S_{f}^{\varphi}$ ($S_{0}^{\phi}$) is the selective phase shift, conditionally changing the phase of target states (zero state) by a factor of $\varphi$ ($\phi$), expressed as ($i=\sqrt{-1}$)
\begin{align}
S_{f}^{\varphi} & =  I-\left(-e^{i\varphi}+1\right)\sum_{x\in f^{-1}\left(1\right)}\left|x\right\rangle \left\langle x\right|, \label{eq:s_varphi_f}\\
S_{0}^{\phi} & =  I-\left(-e^{i\phi}+1\right)\left|0\right\rangle \left\langle 0\right|,
\end{align}
and the phases $\{\phi_{j}, \varphi_{j}: 1\le j\le l\}$  satisfy the following multiphase matching condition \cite{Yoder2014}
\begin{equation}
\phi_{j}=\varphi_{l-j+1}=-2{\rm arccot}\left(\sqrt{1-\gamma^{2}}\tan\left(2\pi j/L\right)\right),\label{eq:multiphase-matching-condition}
\end{equation}
where
$L=2l+1$, $\gamma=T_{1/L}^{-1}(1/\delta)$, $\delta\in(0,1)$ and $T_{L}(x)$ is the $L^{{\rm th}}$ Chebyshev polynomial of the first kind \cite{Mason2002}, defined as
\begin{equation}
T_{L}\left(x\right)=
\left\{ {\begin{array}{*{20}{l}}
\cos\left(L\arccos\left(x\right)\right), & {\rm if}\thinspace\left|x\right|\le1,\\
\cosh\left(L{\rm arcosh}\left(x\right)\right), & {\rm if}\thinspace x\ge1,\\
\left(-1\right)^{L}\cosh\left(L{\rm arcosh}\left(-x\right)\right), & {\rm if}\thinspace x\le-1.
\end{array}} \right.
\label{eq:T_L_x}
\end{equation}

The final state of Yoder's algorithm can be expressed as
\begin{equation}
\left|C_{L}\right\rangle =\sqrt{P_{L}}\left|\alpha\right\rangle +\sqrt{1-P_{L}}\left|\beta\right\rangle ,\label{eq:final-state-C_L}
\end{equation}
where $P_{L}$ denotes the success probability, satisfying 
\begin{equation}
P_{L}=1-\delta^{2}T_{L}^{2}\left[T_{1/L}\left(1/\delta\right)\sqrt{1-\lambda}\right],\label{eq:success_probability}
\end{equation}
and for a given $L$, $P_{L}\ge1-\delta^{2}$ as long as
\begin{equation}
\lambda\ge \bigg(\frac{\ln\left(2/\delta\right)}{L}\bigg)^{2}\equiv \omega. \label{eq:omega}
\end{equation}

Then, for a given $\lambda$, to ensure the success probability no less than $1-\delta^{2}$, based on Eq.~(\ref{eq:omega}), letting $\omega\le\lambda$, the following condition of $L$ can be obtained \cite{Yoder2014}, i.e.,
\begin{equation}
L\ge\frac{\ln\left(2/\delta\right)}{\sqrt{\lambda}}\equiv L_{\rm{min}},
\label{eq:condition-L-Yoder-lam}
\end{equation}
which demonstrates the fixed-point property.
However, in the case of unknown $\lambda$, $L_{\rm{min}}$ is unknown.
For this, it is assumed that there exists a known lower bound $\lambda_{0}$ of $\lambda$ \cite{Yoder2014}.
Then, from Eq.~(\ref{eq:omega}), letting $\omega\le\lambda_{0}$, it follows that
\begin{equation}
L\ge\frac{\ln\left(2/\delta\right)}{\sqrt{\lambda_{0}}}\equiv L_{0}.
\label{eq:condition-L-Yoder-lam_0}
\end{equation}
Note that $L_{0}$
is known, because
$\lambda_0$ is known. Therefore, the query complexity of Yoder's  algorithm is actually in $O(1/\sqrt{\lambda_{0}})$, rather than $O(1/\sqrt{\lambda})$,
which is independent on $\lambda$ and not really optimal.
For example, if the lower bound $\lambda_{0}=1/N$, while $\lambda=M/N=1/\sqrt{N}$ (i.e., $M=\sqrt{N}$), then the order of query complexity of Yoder's algorithm is $O(1/\sqrt{\lambda_{0}})=O(\sqrt{N})$,
which is actually the same as that of classical search, i.e., $O(1/\lambda)=O(\sqrt{N})$,
while the algorithm that achieves the real quadratic speedup over classical algorithms should be in $O(1/\sqrt{\lambda})=O(\sqrt[4]{N})$.

\section{Hybrid fixed-point and trail-and-error quantum search  \label{sec:Hybrid-quantum-search}}

As described in Section~2,
choosing the number of iterations relying on the lower bound $\lambda_{0}$ of $\lambda$, results in the problem that query complexity of Yoder's algorithm \cite{Yoder2014} is not really optimal.
In addition, as described in Ref.~\cite{Okamoto2001},
the existing deterministic trial-and-error algorithm
requires more than 1 times iterations than the randomized version
\cite{Boyer1998}, due to
the success probability of the original Grover algorithm \cite{Grover1996} 
oscillates intensively about $\lambda$.
We expect to handle these problems by
hybridizing the fixed-point method with the trial-and-error method,
that is, we trials Yoder's sequence (defined by Eq.~(\ref{eq:sequence-of-operations})) multiple times and
exponentially increase the number of iterations.
At this time, the condition Eq.~(\ref{eq:condition-L-Yoder-lam}) of $\lambda$, rather than Eq.~(\ref{eq:condition-L-Yoder-lam_0}) of $\lambda_0$, could be satisfied rapidly, and
after that the success probability at each trial would \emph{always} (not just often) be no less than a given value, due to the fixed-point property.
In this way, we could expect that a target item would be found
with the expected Oracle queries in the real optimal order,
and fewer number of iterations than the existing trial-and-error algorithms could be cost.
First, we give the following lemma.

\begin{lemma}
\indent
For any real number $\lambda\in\left(0,1\right)$, $\delta\in\left(0,1\right)$ and integer $L>0$, there exists a lower bound of the success probability $P_{L}$ of Eq.~(\ref{eq:success_probability}), denoted by $P_{L}^{lb}$, which is given as
\begin{equation}
P_{L}^{lb}  \equiv  \begin{cases}
1-\delta^{2}T_{\sqrt{1-L^{2}/L_{cri}^{2}}}^{2}\left(1/\delta\right), & {\rm if}\thinspace L\le L_{cri},\\
1-\delta^{2}, & {\rm if}\thinspace L>L_{cri},
\end{cases}\label{eq:P_L_lb}
\end{equation}
where $T_L(x)$ is defined by Eq.~(\ref{eq:T_L_x}), and
\begin{equation}
L_{cri}=\frac{{\rm arcosh}\left(1/\delta\right)}{{\rm arcosh}\left(1/\sqrt{1-\lambda}\right)}.\label{eq:L_cri}
\end{equation}
\label{lemma:lower-bound-P_L_lb}
\end{lemma}

\emph{\textbf{Proof:}}
First of all, based on Eq.~(\ref{eq:L_cri}), for any $\lambda\in\left(0,1\right)$, $\delta\in\left(0,1\right)$ and $L>0$, $P_{L}$ of Eq.~(\ref{eq:success_probability}) can be rewritten as
\begin{equation}
P_{L}  =  1-\delta^{2}T_{L}^{2}\left[T_{1/L}\left(1/\delta\right)T_{1/L_{cri}}^{-1}\left(1/\delta\right)\right].\label{eq:P_L-expressed-by-L_cri}
\end{equation}
Note that, for any $x>0$ and $\theta\in\left[0,\pi/2\right]$, we can derive the following inequality about the hyperbolic cosine function, i.e.,
\begin{align}
\cosh\left(x\right) & =  \sum_{n=0}^{+\infty}\frac{x^{2n}}{\left(2n\right)!}\left(\cos^{2}\left(\theta\right)+\sin^{2}\left(\theta\right)\right)^{n}\nonumber \\
 & \le  \frac{1}{2}\sum_{n=0}^{\infty}\frac{x^{2n}}{\left(2n\right)!}\left[\left(\cos\theta+\sin\theta\right)^{2n}+\left(\cos\theta-\sin\theta\right)^{2n}\right]\nonumber \\
 & =  \cosh\left(x\sin\theta\right)\cosh\left(x\cos\theta\right).\label{eq:inequality-of-cosh}
\end{align}
Then, in the case of $L\le L_{cri}$, substituting $x=L^{-1}{\rm arcosh}\left(1/\delta\right)$ and $\theta=\arcsin\frac{L}{L_{cri}}$ into Eq.~(\ref{eq:inequality-of-cosh}), we obtain
\begin{equation}
1\le  T_{L}\left[T_{1/L}\left(1/\delta\right)T_{1/L_{cri}}^{-1}\left(1/\delta\right)\right]\le T_{\sqrt{1-L^{2}/L_{cri}^{2}}}\left(1/\delta\right).\label{eq:inequality-of-T_L}
\end{equation}
Furthermore, from Eqs.~(\ref{eq:P_L-expressed-by-L_cri}) and (\ref{eq:inequality-of-T_L}) it follows that
\begin{equation}
P_{L}  \ge 1-\delta^{2}T_{\sqrt{1-L^{2}/L_{cri}^{2}}}^{2}\left(1/\delta\right).\label{eq:P_L-ge-P_L_lb}
\end{equation}
Finally, in the case of $L>L_{cri}$, $P_{L}\ge1-\delta^{2}$ can be easily obtained due to the fact that $\left|T_{L}\left(x\right)\right|\le1$ for $\left|x\right|\le1$. Therefore, the conclusion  is proved. $\hfill\blacksquare$

Now, we are ready to describe the hybrid quantum search algorithm for the unknown $\lambda$ (the corresponding flow diagram is shown in Figure~1):

Step 1: Initialize $k=0$, and let $\delta=0.5659$ and $c=1.523$. (The reason for such selection of $\delta$ and $c$ is given in the proof of Theorem \ref{theorem:result-MMPQS} below.)

Step 2: Increase $k$ by 1.

Step 3: Apply $H$ to the state $\left|0\right\rangle $, and measure the system. If the outcome $\left|x_{0}\right\rangle $ is a target item, i.e., $f\left(x_{0}\right)=1$, then output $x_{0}$ and stop the procedure.

Step 4: Prepare the initial state $\left|\psi\right\rangle =H\left|0\right\rangle $, and perform on $\left|\psi\right\rangle $ the Yoder's sequence
$\prod_{j=1}^{l}G\left(\phi_{j},\varphi_{j}\right)$, where $l=\left\lceil c^{k-1}\right\rceil $  is the number of iterations and
the phases $\left\{ \phi_{j},\varphi_{j}\right\} $ satisfy the multiphase matching condition of Eq.~(\ref{eq:multiphase-matching-condition}).

\begin{center}
\includegraphics[width=5.5cm]{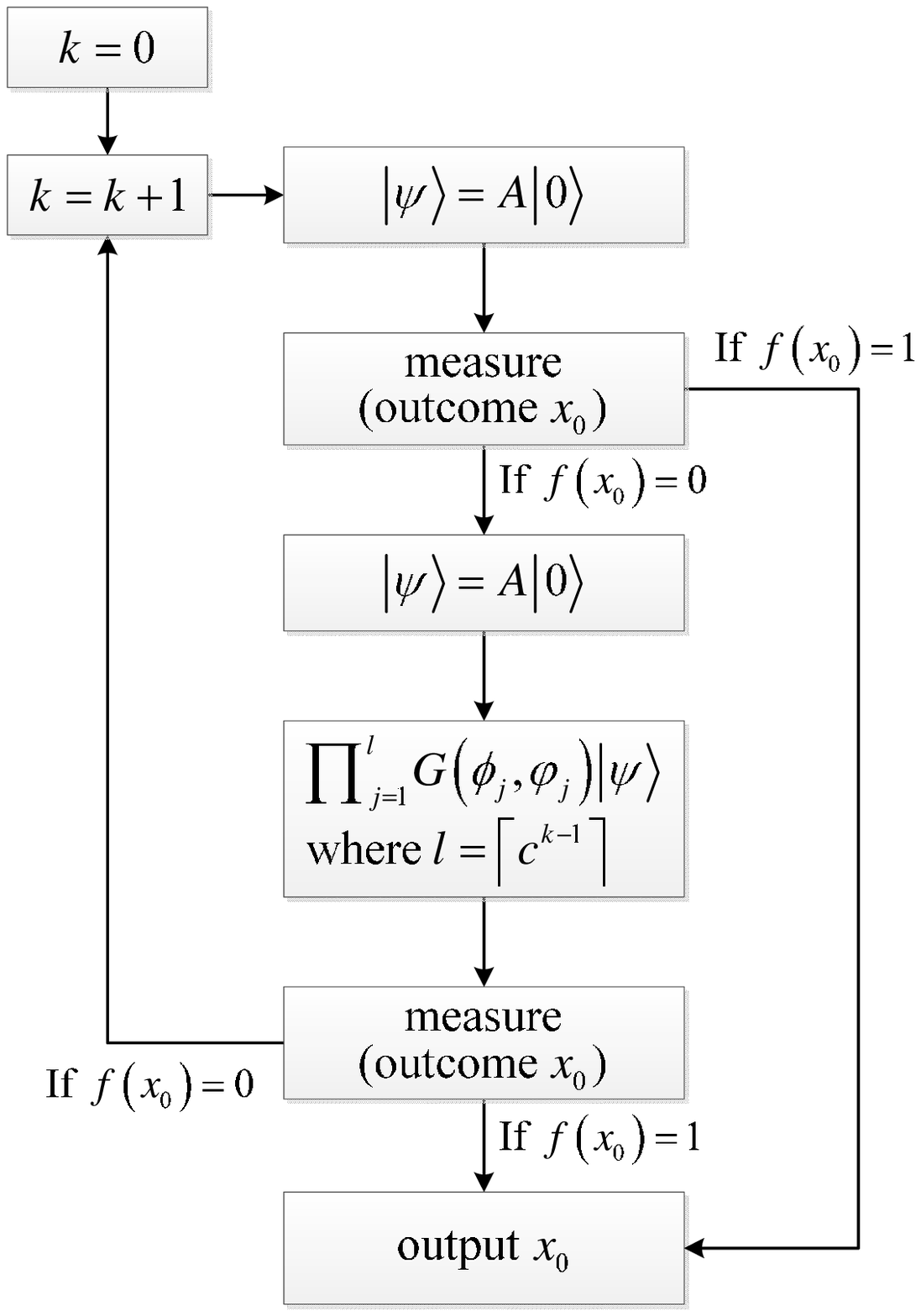}\\[5pt]  
\parbox[c]{15.0cm}{\footnotesize{\bf Fig.~1.} Flow diagram of the hybrid
fixed-point and trail-and-error quantum search algorithm.
\label{fig:flow-diagram-search}}
\end{center}


Step 5: Measure the system. If the outcome $\left|x_{0}\right\rangle $ is a target item, then output $x_{0}$ and stop the procedure. Otherwise, go to Step 2.

\begin{theorem}
This algorithm outputs a target item in expected number of Oracle queries in $O(1/\sqrt{\lambda})$.
\label{theorem:result-MMPQS}
\end{theorem}

\emph{\textbf{Proof:}}
On the one hand, if $\lambda\ge1-\delta^{2}$, then Step 3 ensures that a target item can be measured soon. On the other hand, if $0<\lambda<1-\delta^{2}$, we let
\begin{equation}
l_{cri}=\bigg\lceil \frac{{\rm arcosh}\left(1/\delta\right)}{2{\rm arcosh}\left(1/\sqrt{1-\lambda}\right)}-\frac{1}{2}\bigg\rceil, \label{eq:l_cri}
\end{equation}
and set
\begin{equation}
s_{0}=\left\lfloor \log_{c}l_{cri}\right\rfloor +1.
\label{eq:s_0}
\end{equation}
Then, using $l_{s}$, $Q_{s}$ and $P_{L_{s}}$ to denote the value of the number of iterations, the probability of occurrence and the probability of success on the $s$-th round of the main loop (Steps 2 to 5), the expected number of Oracle queries of the whole
algorithm can be expressed by
\begin{equation}
E\left(T_{all}\right)=E\left(T_{1}\right)+E\left(T_{2}\right),\label{eq:E_T_all-in-improved-search}
\end{equation}
where $E\left(T_{1}\right)$ and $E\left(T_{2}\right)$  denote the expected queries the algorithm takes while $1\le s\le s_{0}$ and $s\ge s_{0}+1$, respectively,
\begin{eqnarray}
E\left(T_{1}\right) & = & \sum_{s=1}^{s_{0}}Q_{s}\left(2l_{s}+2\right),  \label{eq:E_T_1} \\
E\left(T_{2}\right) & = & \sum_{s=s_{0}+1}^{\infty}Q_{s}\left(2l_{s}+2\right), \label{eq:E_T_2}
\end{eqnarray}
with the $l_{s}$ and $Q_{s}$ satisfying 
\begin{eqnarray}
l_{s} & = & \big\lceil c^{s-1}\big\rceil, \label{eq:l_s} \\
Q_{s} & = & \prod_{j=1}^{s-1}\big(1-P_{L_{j}}\big). \label{eq:Q_s}
\end{eqnarray}

By virtue of Lemma \ref{lemma:lower-bound-P_L_lb}, we can get the lower bound of the success probability $P_{L_{s}}$, i.e.,
\begin{align}
P_{L_{s}} &  \ge \begin{cases}
1-\delta^{2}T_{\sqrt{1-c^{-2\left(s_{0}+1-s\right)}}}^{2}\left(1/\delta\right), & {\rm if}\thinspace1\le s\le s_{0},\\
1-\delta^{2}, & {\rm if}\thinspace s\ge s_{0}+1,
\end{cases}\nonumber \\
  & \equiv  P_{L_{s}}^{lb},\label{eq:lower-bound-of-P_L_s}
\end{align}
and further obtain the upper bound of the occurrence probability $Q_{s}$, i.e.,
\begin{align}
Q_{s_{0}+1} &  \le  \delta^{2}T_{\sqrt{1-c^{-2}}}^{2}\left(1/\delta\right)\equiv Q_{s_{0}+1}^{ub},\nonumber \\
Q_{s_{0}+1+u} &  \le Q_{s_{0}+1}^{ub}\delta^{2u},\thinspace u\ge0.\label{eq:upper-bound-of-Q_s}
\end{align}
Then, based on Eqs.~(\ref{eq:s_0},  
\ref{eq:E_T_1}--\ref{eq:upper-bound-of-Q_s}) we can obtain
\begin{align}
E\left(T_{1}\right)  & \le \sum_{s=1}^{s_{0}}\left(2c^{s-1}+4\right)\nonumber \\
 & \le  \frac{2c}{c-1}l_{cri}+4\log_{c}l_{cri}+4, \label{eq:T_1-upper-bound}
\end{align}
and
\begin{align}
E\left(T_{2}\right) &  = \sum_{u=0}^{\infty}Q_{s_{0}+1+u}\left(2l_{s_{0}+1+u}+2\right)\nonumber \\
 &  \le Q_{s_{0}+1}^{ub}\left(\frac{2c}{1-c\delta^{2}}l_{cri}+\frac{4}{1-\delta^{2}}\right),\label{eq:T_2-upper-bound}
\end{align}
where we have assumed that $c<\delta^{-2}$. Finally, from Eqs.~(\ref{eq:E_T_all-in-improved-search}, \ref{eq:upper-bound-of-Q_s}-\ref{eq:T_2-upper-bound}) and $l_{cri}\gg\log_{c}l_{cri}\gg1$ for $\lambda\ll1-\delta^{2}$, it follows that
\begin{eqnarray}
E\left(T_{all}\right)  & \le&  \bigg(\frac{2c}{c-1}+\frac{2Q_{s_{0}+1}^{ub}c}{1-c\delta^{2}}\bigg)l_{cri}+4\Big(\log_{c}l_{cri}+\frac{Q_{s_{0}+1}^{ub}}{1-\delta^{2}}+1\Big) \nonumber \\
  &\approx & \Big(\frac{c}{c-1}+\frac{c\delta^{2}T_{\sqrt{1-c^{-2}}}^{2}\left(1/\delta\right)}{1-c\delta^{2}}\Big)\frac{{\rm arcosh}\left(1/\delta\right)}{\sqrt{\lambda}} \nonumber \\
  & \equiv & g\left(\delta,c\right)/\sqrt{\lambda}, \label{eq:T_all_up}
\end{eqnarray}
and numerical calculation shows that
\begin{align}
   & \min\left\{ g\left(\delta,c\right):0<\delta<1,1<c<\delta^{-2}\right\} \nonumber \\
  =\; & g\left(\delta\approx0.5659,c\approx1.523\right)\nonumber \\
  \approx\; & 5.643.
\end{align}
Therefore, the query complexity of our algorithm is in $O(1/\sqrt{\lambda})$.
In addition, we can see that $\delta=0.5659$ and $c=1.523$ are just the optimal parameters minimizing the upper bound of the expected query complexity. $\hfill\blacksquare$

\section{Discussion and conclusion\label{sec:Discussions_Conclusion}}
In this section, we first discuss the comparisons between our algorithm and the 
existing quantum search algorithms \cite{Boyer1998,Okamoto2001,Grover2005,Younes2008,Younes2013,Yoder2014}  in the case of unknown $\lambda$ and then give a brief conclusion.
Table~1 lists the method, query complexity,
and phase(s) of our algorithm and
other algorithms.
The main advantages of our algorithm are discussed  in detail  as follows.

\begin{center}
\begin{threeparttable}
\footnotesize {\bf Table 1.} Detailed comparisons between our algorithm and other algorithms. \\
\vspace{2mm}
\begin{tabular}{>{\centering}m{2.8cm}>{\centering}m{6.5cm}>{\centering}m{3cm}>{\centering}m{2.8cm}}
\hline
Algorithm & Method & Query complexity  & Phase(s) \tabularnewline \hline
Grover \cite{Grover2005} & fixed-point   & $O\big(\frac{\ln\left(1/\delta\right)}{\lambda_{0}}\big)$ \tnote{$^{\rm a,b}$} & $\pi/3$ \tabularnewline
Yoder et al. \cite{Yoder2014} & fixed-point &
$O\big(\frac{\ln\left(2/\delta\right)}{\sqrt{\lambda_{0}}}\big)$ \tnote{$^{\rm a,b}$}  &  $\{\phi_{j}, \varphi_{j}\}$ of
Eq.~(\ref{eq:multiphase-matching-condition}) \tabularnewline
Boyer et al. \cite{Boyer1998} & trail-and-error (randomized)  &
$\le4/\sqrt{\lambda}$ & $\pi$ \tabularnewline
Younes et al. \cite{Younes2008} & trail-and-error (randomized) \tnote{$^{\rm c}$}  & $\le4\sqrt{2}/\sqrt{\lambda}$  &  $\pi$ \tabularnewline
Younes \cite{Younes2013} & trail-and-error (randomized) \tnote{$^{\rm d}$} & $\le61.42/\sqrt{\lambda}$ \tnote{$^{\rm b}$}  & $1.91684\pi$ \tabularnewline
Okamoto et al. \cite{Okamoto2001} & trail-and-error (deterministic) & $\le8.378/\sqrt{\lambda}$ & $\pi$  \tabularnewline
Pro. Alg. & fixed-point + trail-and-error (deterministic) & $\le5.643/\sqrt{\lambda}$ \tnote{$^{\rm b}$} & $\{\phi_{j}, \varphi_{j}\}$ of
Eq.~(\ref{eq:multiphase-matching-condition}) \tabularnewline
\hline
\end{tabular}
\begin{tablenotes}
\item[$^{\rm a}$] $\lambda_0$ is the assumed known lower bound of the unknown fraction $\lambda$ of target items in Refs.~\cite{Grover2005,Yoder2014}, and the obtained success probability is no less than $1-\delta^2$, where $\delta\in(0,1)$.
\item[$^{\rm b}$] The number of Grover iterations is half of it. Note that
in the derivation of query complexity,
the viewpoint of Ref.~\cite{Yoder2014} has been adopted, that is, one Grover iteration with arbitrary phases other than $\pi$ contains two queries of the standard quantum Oracle with phase-$\pi$, defined by Eq.~{(\ref{eq:S_pi_f})}.
\item[$^{\rm c}$] Different from Boyer's algorithm, the internal Grover's algorithm is replaced by the partial diffusion algorithm \cite{Younes2004}.
\item[$^{\rm d}$] The internal Grover's algorithm is replaced by the ($1.91684\pi$) fixed-phase  algorithm \cite{Younes2013}.
\end{tablenotes}
\end{threeparttable}
\end{center}

Compared with the fixed-point quantum search algorithms \cite{Grover2005,Yoder2014}, as shown in Eq.~(\ref{eq:T_all_up}) our algorithm indeed reaches the optimal scaling for quantum search, i.e., $O(1/\sqrt{\lambda})$, while, Refs.~\cite{Grover2005,Yoder2014} do not. Reasons are in the following:
First, for the original fixed-point quantum search algorithm \cite{Grover2005},
to achieve a success probability no less than $1-\delta^2$ ($\delta\in(0,1)$) for a unknown $\lambda$ with a known lower bound $\lambda_0$,
the number of required Oracle queries  \cite{Tulsi2006} is in $O\big(\frac{\ln\left(1/\delta\right)}{\lambda_{0}}\big)\ge O\big(\frac{\ln\left(1/\delta\right)}{\lambda}\big)$,
yielding the loss of quadratic speedup over classical exhaustive search, which is in $O(1/\lambda)$.
Second, as shown in Eq.~(\ref{eq:condition-L-Yoder-lam_0}), the query complexity of Yoder's algorithm  \cite{Yoder2014} is in $O\big(\frac{\ln\left(2/\delta\right)}{\sqrt{\lambda_{0}}}\big)$, which indeed achieves a quadratic speedup over
the original fixed-point algorithm.
However, for any $\lambda\ge\lambda_0$,
the same as Ref.~\cite{Grover2005},
the least number of iterations of
Yoder's algorithm
is fixed.
For example, if $\lambda_{0}=1/N$, while $\lambda=(N-1)/N\approx 1$, then
a target item could almost be found by performing the classical search once, but the complexity of Yoder's algorithm is still in $O(1/\sqrt{\lambda_{0}})=O(\sqrt{N})$.
Therefore, Yoder's algorithm
does not
achieve the quadratic speedup over classical search. Note that,
this advantage of our algorithm comes from the use of trial-and-error method.

Compared with the trial-and-error quantum search algorithms \cite{Boyer1998,Younes2008,Younes2013,Okamoto2001},
first, with respect to the method, our algorithm
does not contain randomness (i.e., random selection of the number of iterations), and enables the success probability \emph{always} no less than an arbitrary given lower bound between 0 and 1 after the number of iterations exceeds the critical value (i.e., $l_{cri}$ defined by  Eq.~(\ref{eq:l_cri})).
While, the existing randomized trial-and-error algorithms \cite{Boyer1998,Younes2008,Younes2013}
rely on the randomness, and the existing deterministic trial-and-error algorithm \cite{Okamoto2001} can only makes the success probability \emph{often} (but not always) no less than a lower bound, fixed at 3/4.
Second, with respect to the query/iteration complexity, compared with the existing deterministic trial-and-error algorithm \cite{Okamoto2001},
our query complexity can be reduced by $(8.378-5.643) / 8.378 \approx 1/3 $.
Although, the upper bound of query complexity of our algorithm displayed in Table 1 is slightly higher than Boyer's randomized
algorithm \cite{Boyer1998} (about
$5.643/4 \approx 1.4$ times), however,
in the derivation of query complexity,
we have adopted
the viewpoint in Ref.~\cite{Yoder2014} that
one Grover iteration with arbitrary phases
queries two standard quantum Oracle (denoted by $S^\pi_{f}$) with phase-$\pi$,
which flips the ancilla qubit when the input is the target state, i.e.,
\begin{equation}
S^\pi_{f}\left|x\right\rangle \left|y\right\rangle =\left|x\right\rangle \left|y\oplus f\left(x\right)\right\rangle.
\label{eq:S_pi_f}
\end{equation}
Therefore, in fact,
our algorithm has fewer number of iterations, about
$1-\frac{5.643}{2}/4\approx 30\%$ can be saved than Ref.~\cite{Boyer1998}.
This advantage comes from the use of fixed-point method in our algorithm.

Finally, it is worth mentioning that,
for the quantum Oracle with arbitrary phases (i.e., $S^\varphi_f$, defined by Eq.~(\ref{eq:s_varphi_f})),
Yoder et al. designed a feasible quantum circuit,
including two standard quantum Oracle $S^\pi_f$ with phase-$\pi$,
which appears to show that
the complexity of $S^\varphi_f$ is twice that of $S^\pi_f$.
However, we have noticed that
in many practical applications of quantum search, e.g., key search in classical cryptography \cite{Grassl2016,Kim2018,Denisenko2019}, the satisfiability problem \cite{Cheng2007,Yang2009},
quantum machine learning \cite{Lee2016}, etc.,
the design of the standard quantum Oracle $S^\pi_f$
contains a
quantization (denoted by $Q_f$) of the classical Oracle
and the corresponding uncomputation step
(denoted by $Q^\dagger_f$) which restores the ancillary qubits to their initial state \cite{Matuschak2019}, as shown in Fig.~2.
Therefore, the complexity of $S^\pi_f$ is twice that of $Q_f$, also as pointed in Ref.~\cite{Chailloux2017}.
What's more,
based on this, we can find out that
follow the circuit of $S^\varphi_f$  in Ref.~\cite{Yoder2014}, as depicted in Fig.~2,
the $Q^\dagger_f$ in the first $S^\pi_f$ and the $Q_f$ in the second $S^\pi_f$ could be omitted because they offset each other ($Q^\dagger_f Q_f=I$).
Therefore,
the complexity of quantum Oracle $S^\varphi_f$ with arbitrary phases would also be twice that of $Q_f$, which is the same as that of the standard Oracle $S^\pi_f$.
In this case,
our algorithm will has
the least query complexity among the current algorithms for the unknown $\lambda$ \cite{Boyer1998,Grover2005,Yoder2014,Younes2008,Younes2013,Okamoto2001}.
While, this conclusion may be related to the specific application scenario of the algorithms,
and more rigorous proofs can be studied further in the future.
\begin{center}
\includegraphics[width=11.5cm]{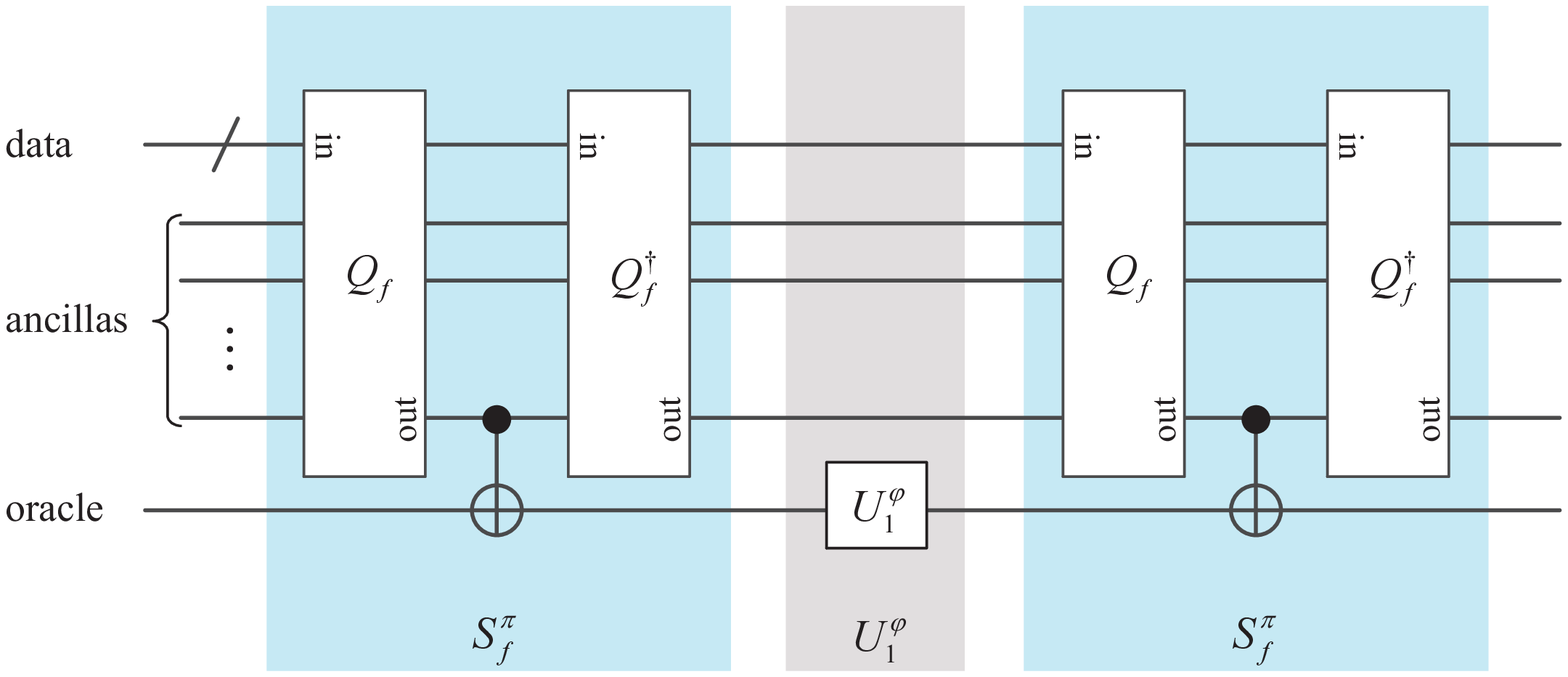}\\[5pt]  
\parbox[c]{15.0cm}{\footnotesize{\bf Fig.~2.} (color online) Circuit for the quantum Oracle $S^\varphi_f$ with arbitrary phases, defined by Eq.~(\ref{eq:s_varphi_f}), where $S^\pi_f$ denotes the quantum Oracle with phase-$\pi$ (defined by Eq.~(\ref{eq:S_pi_f})), $Q_f$ is the quantization of the classical Oracle, $Q^\dagger_f$ corresponds to the uncomputation step to restore the ancillary qubits, and $U_1^\varphi$ is the quantum gate adding a phase of $e^{i\varphi}$ to $\left|1\right>$ state.
\label{fig:Circuit_Oracle_arbitrary_phases}}
\end{center}

In summary, we have presented the first quantum search algorithm hybridizing the fixed-point method with the trial-and-error method for the unknown $\lambda$,
which integrates the advantages of both methods and solves
the non-optimal problem of the ``optimal'' fixed-point algorithm \cite{Yoder2014} and
the problem that
the deterministic trial-and-error algorithm \cite{Okamoto2001}
requires more than 1 times queries or iterations
than the randomized versions.
In our algorithm,
the Yoder's sequence \cite{Yoder2014} is conducted multiple trials,
and the number of iterations increases exponentially  along with the number of trials.
The upper bound of expected queries of our algorithm to find a target item was derived,
as well as the optimal parameters,
which minimize the query complexity.
Our study reconfirms the practicality of the fixed-point method,
and shows that it is very advantageous to hybridize the trial-and-error method and the fixed-point method, which provides a different approach for the research on quantum search algorithms and can be applied to a variety of scenarios where Grover's search is used \cite{Grassl2016,Kim2018,Denisenko2019,Cheng2007,Yang2009,Lee2016}.


\addcontentsline{toc}{chapter}{Acknowledgment}
\section*{Acknowledgment}
We are grateful to the reviewers for their constructive comments to improve this paper, and we thank He-Liang Huang, Feng-Guang Li, Fan Liu and Ao-Di Liu for useful discussions.


\addcontentsline{toc}{chapter}{References}
\begin{thebibliography}{99}\footnotesize
\itemsep=-3pt plus.2pt minus.2pt   

\bibitem {Grover1996} Grover L K
    {1996 \emph{Proceedings of the Twenty-eighth Annual ACM Symposium on Theory of Computing}, New York, pp. 212--219}

\bibitem {Grover1997} Grover L K
    \href {http://doi.org/10.1103/PhysRevLett.79.325}
    {1997 \emph{Phys. Rev. Lett.} \textbf{79} 325}

\bibitem {Biron1999} Biron D, Biham O, Biham E, Grassl M and Lidar D A
    {1999 \emph{Quantum Computing and Quantum Communications} (Berlin: Springer) pp. 140--147}

\bibitem {Long1999} Long G L, Li Y S, Zhang W L and Niu L
    \href {http://doi.org/10.1016/S0375-9601(99)00631-3}
    {1999 \emph{Phys. Lett. A} \textbf{262} 27}

\bibitem {Long2001} Long G L
    \href {http://doi.org/10.1103/PhysRevA.64.022307}
    {2001 \emph{Phys. Rev. A} \textbf{64} 022307}

\bibitem {Long2002} Long G L, Li X and Sun Y
    \href {http://doi.org/10.1016/S0375-9601(02)00055-5}
    {2002 \emph{Phys. Lett. A} \textbf{294} 143}

\bibitem {Li2007} Li P C and Li S Y
    \href {http://doi.org/10.1016/j.physleta.2007.02.029}
    {2007 \emph{Phys. Lett. A} \textbf{366} 42}


\bibitem {Zhang2011} Zhang Y Y, Hu H P and Lu S F
\href {http://doi.org/10.1088/1674-1056/20/4/040309}
{2011 \emph{Chinese Phys. B} \textbf{20} 040309}

\bibitem {Sun2012} Sun J, Lu S F, Liu F and Yang L P
\href {http://doi.org/10.1088/1674-1056/21/1/010306}
{2012 \emph{Chinese Phys. B} \textbf{21} 010306}

\bibitem {Li2018a} Li F G, Bao W S, Zhang S, Wang X, Huang H L, Li T and Ma B W
    \href {http://doi.org/10.1088/1674-1056/27/1/010308}
    {2018 \emph{Chinese Phys. B} \textbf{27} 10308}

\bibitem {Toyama2008} Toyama F M, Dijk W V, Nogami Y, Tabuchi M and Kimura Y
    \href {http://doi.org/10.1103/PhysRevA.77.042324}
    {2008 \emph{Phys. Rev. A} \textbf{77} 042324}

\bibitem {Toyama2009} Toyama F M, Kasai S, Dijk W V and Nogami Y,
    \href {http://doi.org/10.1103/PhysRevA.79.014301}
    {2009 \emph{Phys. Rev. A} \textbf{79} 014301}

\bibitem {Toyama2013} Toyama F M, Dijk W V and Nogami Y
    \href {http://doi.org/10.1007/s11128-012-0498-0}
    {2013 \emph{Quantum Inf. Process.} \textbf{12} 1897}

\bibitem {Toyama2019} Toyama F M and Dijk W V
    \href {http://doi.org/10.1139/cjp-2018-0452}
    {2019 \emph{Can. J. Phys.} \textbf{97} 777}

\bibitem {Zhong2009} Zhong P C, Bao W S and Wei Y
    \href {http://doi.org/10.1088/0256-307X/26/2/020301}
    {2009 \emph{Chinese Phys. Lett.} \textbf{26} 020301}

\bibitem {Giri2017} Giri P R and Korepin V E
    \href {http://doi.org/10.1007/s11128-017-1768-7}
    {2017 \emph{Quantum Inf. Process.} \textbf{16} 315}

\bibitem {Zhang2018} Zhang K and Korepin V E
    \href {http://doi.org/10.1007/s11128-018-1907-9}
    {2018 \emph{Quantum Inf. Process.} \textbf{17} 143}

\bibitem {Mehri-Dehnavi2018} Mehri-Dehnavi H, Dashtianeh H, Kuchaksaraei H Y, Khoshdareh M M,  Movahhedian H and Rahimi R
    \href {http://doi.org/10.1007/s10773-018-3880-6}
    {2018 \emph{Int. J. Theor. Phys.} \textbf{57} 3668}

\bibitem {Byrnes2018} Byrnes T, Forster G and Tessler L
    \href {http://doi.org/10.1103/PhysRevLett.120.060501}
    {2018 \emph{Phys. Rev. Lett.} \textbf{120} 060501}

\bibitem {Ma2017} Ma B W, Bao W S, Li T, Li F G, Zhang S, and Fu X Q
\href {http://doi.org/10.1088/0256-307X/34/7/070305}
{2017 \emph{Chin. Phys. Lett.} \textbf{34} 070305}

\bibitem {Li2014} Li T, Bao W S, Lin W Q, Zhang H and Fu X Q
    \href {http://doi.org/10.1088/0256-307X/31/5/050301}
    {2014 \emph{Chinese Phys. Lett.} \textbf{31} 050301}

\bibitem {Li2018} Li T, Bao W S, Huang H L, Li F G, Fu X Q, Zhang S, Guo C, Du Y T, Wang X and  Lin J
    \href {http://doi.org/10.1103/PhysRevA.98.062308}
    {2018 \emph{Phys. Rev. A} \textbf{98} 062308}

\bibitem {Li2019} Li T, Fu X Q, Wang Y, Zhang S, Wang X, Du Y T and Bao W S
    \href {http://arxiv.org/abs/1908.00269}
    {2019 \emph{arXiv}: 1908.00269 [quant-ph]}

\bibitem {Pan2019}
Pan M H and Qiu D W
\href {http://doi.org/10.1103/PhysRevA.100.012349}
{2019 \emph{Phys. Rev. A} \textbf{100} 012349}

\bibitem {Pan2019a}
Pan M H, Qiu D W, Mateus P and Gruska J
\href {http://doi.org/10.1016/j.tcs.2018.10.001}
{2019 \emph{Theor. Comput. Sci.} \textbf{773} 138}

\bibitem {Bennett1997} Bennett C H, Bernstein E, Brassard G and Vazirani U
    \href {http://doi.org/10.1137/S0097539796300933}
    {1997 \emph{SIAM J. Comput.} \textbf{26} 1510}

\bibitem {Boyer1998} Boyer M, Brassard G, H{\o}yer P and Tapp A
    \href {http://doi.org/10.1002/(SICI)1521-3978(199806)46:4/5<493::AID-PROP493>3.0.CO;2-P}
    {1998 \emph{Fortschr. Phys.} \textbf{46} 493}

\bibitem {Zalka1999} Zalka C
    \href {http://doi.org/10.1103/PhysRevA.60.2746}
    {1999 \emph{Phys. Rev. A} \textbf{60} 2746}

\bibitem {Nielson2000} Nielson M A and Chuang I L
    2000 \emph{Quantum Computation and Quantum Information} (1st edn.) (Cambridge: Cambridge University Press) pp. 269--271

\bibitem {Grover2005a} Grover L K and Radhakrishnan J
{2005 \emph{Proceedings of the Seventeenth Annual ACM Symposium on Parallelism in Algorithms and Architectures} New York, pp. 186--194}

\bibitem {Brassard1997} Brassard G
    \href {http://doi.org/10.1126/science.275.5300.627}
    {1997 \emph{Science} \textbf{275} 627}

\bibitem {Grover2005} Grover L K
    \href {http://doi.org/10.1103/PhysRevLett.95.150501}
    {2005 \emph{Phys. Rev. Lett.} \textbf{95} 150501}

\bibitem {Yoder2014} Yoder T J, Low G H and Chuang I L
    \href {http://doi.org/10.1103/PhysRevLett.113.210501}
    {2014 \emph{Phys. Rev. Lett.} \textbf{113} 210501}

\bibitem {Chakraborty2005} Chakraborty S, Radhakrishnan J and Raghunathan N
    {2005 \emph{Approximation, Randomization and Combinatorial Optimization. Algorithms and Techniques} (Berlin: Springer) pp. 245--256}

\bibitem {Tulsi2006} Tulsi T, Grover L K and Patel A
    \href {http://www.rintonpress.com/journals/qiconline.html#v6n6}
    {2006 \emph{Quantum Inf. Comput.} \textbf{6} 483}

\bibitem {Bhole2016} Bhole G, Anjusha V S and Mahesh T S
    \href {http://doi.org/10.1103/PhysRevA.93.042339}
    {2016 \emph{Phys. Rev. A} \textbf{93} 042339}

\bibitem {Gurnani2017} Gurnani K, Behera B K and Panigrahi P K
    \href {http://arxiv.org/abs/1712.10231}
    {2017 \emph{arXiv}: 1712.10231 [quant-ph]}

\bibitem {Dalzell2017} Dalzell A M, Yoder T J and Chuang I L
    \href {http://doi.org/10.1103/PhysRevA.95.012311}
    {2017 \emph{Phys. Rev. A} \textbf{95} 012311}


\bibitem {Qiu2018} Qiu D W and Zheng S G
    \href {http://doi.org/10.1103/PhysRevA.97.062331}
    {2018 \emph{Phys. Rev. A} \textbf{97} 062331}

\bibitem {Cai2018} Cai G Y and Qiu D W
\href {http://doi.org/10.1016/j.jcss.2018.05.001}
{2018 \emph{J. Comput. Syst. Sci.} \textbf{97} 83}

\bibitem {Younes2008} Younes A, Rowe J and Miller J
\href {http://doi.org/10.1016/j.physd.2007.12.005}
{2008 \emph{Physica D} \textbf{237} 1074}

\bibitem {Younes2013} Younes A
 \href {http://www.naturalspublishing.com/Article.asp?ArtcID=1264}
{2013 \emph{Appl. Math. Inf. Sci.} \textbf{7} 93}

\bibitem {Okamoto2001} Okamoto K and Watanabe O
\href {http://doi.org/10.1007/3-540-44679-6_55}
{2001 \emph{Computing and Combinatorics} (Berlin: Springer) pp. 493--501}

\bibitem {Younes2004} Younes A, Rowe J and Miller J
    \href {http://doi.org/10.1063/1.1834408}
    {2004 \emph{AIP Conf. Proc.} \textbf{734} 171}

\bibitem {Brassard2002} Brassard G, H{\o}yer P, Mosca M and Tapp A
    {2002 \emph{Quantum Computation and Information} (Providence: American Mathematical Society) pp. 53--74}

\bibitem {Mason2002} Mason J C and Handscomb D C
    2002 \emph{Chebyshev Polynomials} (Boca Raton: CRC Press) pp. 2--3

\bibitem {Grassl2016} Grassl M, Langenberg B, Roetteler M and Steinwandt R
\href {http://link.springer.com/10.1007/978-3-319-29360-8_3}
{2016 \emph{Post-Quantum Cryptography} (Switzerland: Springer) pp. 29--43}

\bibitem {Kim2018} Kim P, Han D and Jeong K C
\href {http://doi.org/10.1007/s11128-018-2107-3}
{2018 \emph{Quantum Inf. Process.} \textbf{17} 339}

\bibitem {Denisenko2019} Denisenko D V and Nikitenkova M V
\href {http://doi.org/10.1134/S1063776118120142}
{2019 \emph{J. Exp. Theor. Phys.} \textbf{128} 25--44}

\bibitem {Cheng2007} Cheng S T and Tao M H
\href {http://doi.org/10.1134/S1063776118120142}
{2007 \emph{J. Comput. Syst. Sci.} \textbf{73} 123--136}

\bibitem {Yang2009} Yang W L, Wei H, Zhou F, Chang W L and Feng M
\href {http://doi.org/10.1088/0953-4075/42/14/145503}
{2009 \emph{J. Phys. B-At. Mol. Opt. Phys.} \textbf{42} 145503}

\bibitem {Lee2016} Lee B and Perkowski M
\href {http://doi.org/10.1109/DSD.2016.30}
{2016 \emph{Euromicro Conference on Digital System Design (DSD)} (Cyprus: IEEE) pp. 413--422}

\bibitem {Matuschak2019} Matuschak A and Nielsen M A
\href {https://quantum.country/search}
{https://quantum.country/search [Accessed: 15 September 2019]}

\bibitem {Chailloux2017} Chailloux A, Naya-Plasencia M and Schrottenloher A
{2017 \emph{Advances in Cryptology ¨C ASIACRYPT 2017} (Cham: Springer) pp. 211--240}

\end{thebibliography}

\end{document}